\documentclass[conference]{IEEEtran}
\IEEEoverridecommandlockouts
\usepackage{cite}
\usepackage{amsmath,amssymb,amsfonts}
\usepackage{algorithmic}
\usepackage{graphicx}
\usepackage{textcomp}
\usepackage{xcolor}
\usepackage{subfigure}
\usepackage{adjustbox}
\usepackage{array}
\usepackage{url}
\graphicspath{ {../Figures} }

\def\BibTeX{{\rm B\kern-.05em{\sc i\kern-.025em b}\kern-.08em
    T\kern-.1667em\lower.7ex\hbox{E}\kern-.125emX}}
\begin{document}

\title{An Extended Model for Ecological Robustness to Capture Power System Resilience}




\author{\IEEEauthorblockN{ Hao Huang\IEEEauthorrefmark{1}\IEEEauthorrefmark{2}, \emph{Member, IEEE,} Katherine R. Davis, \emph{Senior Member, IEEE}\IEEEauthorrefmark{2}, H. Vincent Poor, \emph{Fellow, IEEE}\IEEEauthorrefmark{1}}
\IEEEauthorblockA{
\IEEEauthorrefmark{1}Department of Electrical and Computer Engineering, Princeton University, Princeton, NJ 08544, USA \\
\IEEEauthorrefmark{2}Department of Electrical and Computer Engineering, Texas A\&M University, College Station, TX 77843, USA \\
Email:  hh6219@princeton.edu, katedavis@tamu.edu, poor@princeton.edu}
} 


\maketitle
\begin{abstract}
The long-term resilient property of ecosystems has been quantified as ecological robustness (R\textsubscript{ECO}) in terms of the energy transfer over food webs. The R\textsubscript{ECO} of resilient ecosystems favors a balance of food webs' network efficiency and redundancy. By integrating R\textsubscript{ECO} with power system constraints, the authors are able to optimize power systems’ inherent resilience as ecosystems through network design and system operation. A previous model used on real power flows and aggregated redundant components for a rigorous mapping between ecosystems and power systems. However, the reactive power flows also determine power systems resilience; and the power components' redundancy is part of the global network redundancy. These characteristics should be considered for R\textsubscript{ECO}-oriented evaluation and optimization for power systems. 
Thus, this paper extends the model for quantifying R\textsubscript{ECO} in power systems using real, reactive, and apparent power flows with the consideration of redundant placement of generators. Recalling the performance of R\textsubscript{ECO}-oriented optimal power flows under \textit{N-x} contingencies, the analyses suggest reactive power flows and redundant components should be included for R\textsubscript{ECO} to capture power systems’ inherent resilience. 

\end{abstract}

\begin{IEEEkeywords}
Resilience, Power System Resilience, Ecosystems, Biological System Modeling
\end{IEEEkeywords}

\section{Introduction}


Power systems are the backbone for modern society which generate, transfer, and deliver electric energy from different energy resources to end-users. However, the increasingly frequent unexpected events, including cyber attacks and natural disasters, have interrupted power systems’ functionality and compromised its reliability, and thus threaten the security and safety of the whole social welfare. Enhancing the inherent resilience of power systems against any source of disturbances to maintain security and reliability is an urgent task.

Power system resilience has been defined as the ability to avoid or withstand grid stress events without suffering operational compromise or to adapt to and compensate for the resulting strains so as to minimize compromise via graceful degradation \cite{taft2017electric}. Various measures and metrics have been proposed to quantify resilience \cite{panteli2017metrics, clark2017cyber, venkataramanan2020cp}. In \cite{panteli2017metrics}, Panteli \textit{et al.} proposed a set of metrics to quantify power system resilience against natural disasters based on a multi-phase resilience trapezoid. In \cite{clark2017cyber}, Clark and Zonouz proposed a cyber-physical resilience metric using Markov decision process to assess the operational impacts from cyber attacks. Similarly, Tushar \textit{et al.} utilized the graphic analysis on both physical and cyber networks to quantify the cyber-physical transmission resilience as the ability of supplying energy under extreme events \cite{venkataramanan2020cp}. The commonality of above approaches is that they focused on power systems functionality and security against specific types of hazards while neglected the system's inherent ability of absorbing agnostic disturbances.

This meaning of \textit{resilience} can date back to the 1970s, where \textit{C.S. Holling} defines \textit{resilience} in ecology as a measure of the ability to absorb changes of variables and parameters in systems \cite{holling1973resilience}. Ecosystems have evolved over millions of years to survive disturbances. In ecosystems, food webs include the food chains in a graphical representation among different species based on their prey-predator relationships. The graph captures the flow of energy and materials in ecosystems, which contains their resilient structural properties \cite{cohen1993improving}. 
Ecologists have developed methodologies and metrics to capture the resilient traits, such as ecological robustness (R\textsubscript{ECO}), cycling index, and degree of system order (DoSO) \cite{fath2007ecological, ulanowicz2009quantifying}. To mimic the resilient nature from ecosystems, such metrics have been used to guide the resilient design for various human-made networks, including water distribution networks \cite{Tirth2020JCP}, supply chains \cite{Chatterjee_supplychain2020RESS}, and industrial by-product networks \cite{Layton2015ecological}.


Among the above metrics, R\textsubscript{ECO} is directly related to the long-term resiliency in ecosystems \cite{ulanowicz2009quantifying}. It derives from an information theoretic approach considering energy transitions among all species over food webs, representing a function of the pathway \textit{efficiency} and \textit{redundancy}. The value of R\textsubscript{ECO} for robust food webs falls into a specific range, called \textit{Window of Vitality}, which circumscribes sustainable and resilient behavior in ecosystems. 
With a one-to-one mapping between power systems and ecosystems in which buses and generators are modeled as species (actors), real power flows are equated as energy transfer, and power flow direction represents the \textit{prey-predator} relationships, we are able to quantify R\textsubscript{ECO} for power systems. It allows the integration of R\textsubscript{ECO} with power system constraints to optimize power network structure and system operation for enhanced inherent resilience. \textcolor{black}{The improvement of R\textsubscript{ECO} results in fewer operational violations and unsolved contingencies when the system is subject to multiple hazards \cite{huangreco,huangreco2022}.}

However, the previous works on R\textsubscript{ECO} don't consider the reactive power flows as there is no corresponding element in ecosystems. The reactive power flows are related to voltage stability in power systems. Along with the reactive power flows are the shunt capacitors, which can either generate or consume reactive power to regulate voltage for supporting the system. Another distinction between power systems and ecosystems is the redundant components/actors. In power systems, one bus can connect multiple generators with separate controls; while in ecosystems, if the actors are the same species, their energy transfer is aggregated. The component-to-actor based mapping does not capture the redundancy in power systems. Thus, the research question raised in this paper is \textit{How power systems' distinct characteristics will influence the quantification of R\textsubscript{ECO}?} 


To investigate this question, this paper extends the model of mapping power system with ecosystem and calculates R\textsubscript{ECO} with different flow types, namely real, reactive, and apparent power flow, considering two ways of dealing with redundant components in power systems, respectively. This extended model is applied two power system cases, the IEEE 24 Bus RTS System \cite{zimmerman2010matpower} and the Reduced Great Britain (GB) Network \cite{bukhsh2013network}, with different optimal power flow algorithms from \cite{huangreco}. With the \textit{N-x} contingency analyses, it has been observed that the reactive power based R\textsubscript{ECO} is better to capture power systems' inherent resilience; 
and the component redundancy can contribute to the improvement on R\textsubscript{ECO}. 
The main contributions of this paper are as follow:
\begin{itemize}

\item A detailed mapping between power systems and ecosystems is proposed to investigate the application of ecological metrics in power systems studies considering reactive power, apparent power, and component redundancy. This extended R\textsubscript{ECO} evaluation has been implemented in EasySimAuto (ESA) \cite{ESA}.

\item We perform comprehensive analyses of R\textsubscript{ECO} on two power system cases under different optimal power flow algorithms. With the performance under \textit{N-x} contingencies, the results suggest that reactive power flows and redundant components should be included for calculating R\textsubscript{ECO} to better capture power systems’ inherent resilience.


\end{itemize}










\section{Background of Ecological Robustness}
\label{background}

\begin{figure}[h!]
\centering
\includegraphics[trim={1.2mm 1.2mm 1.2mm 1.2mm}, clip,width=0.98\linewidth]{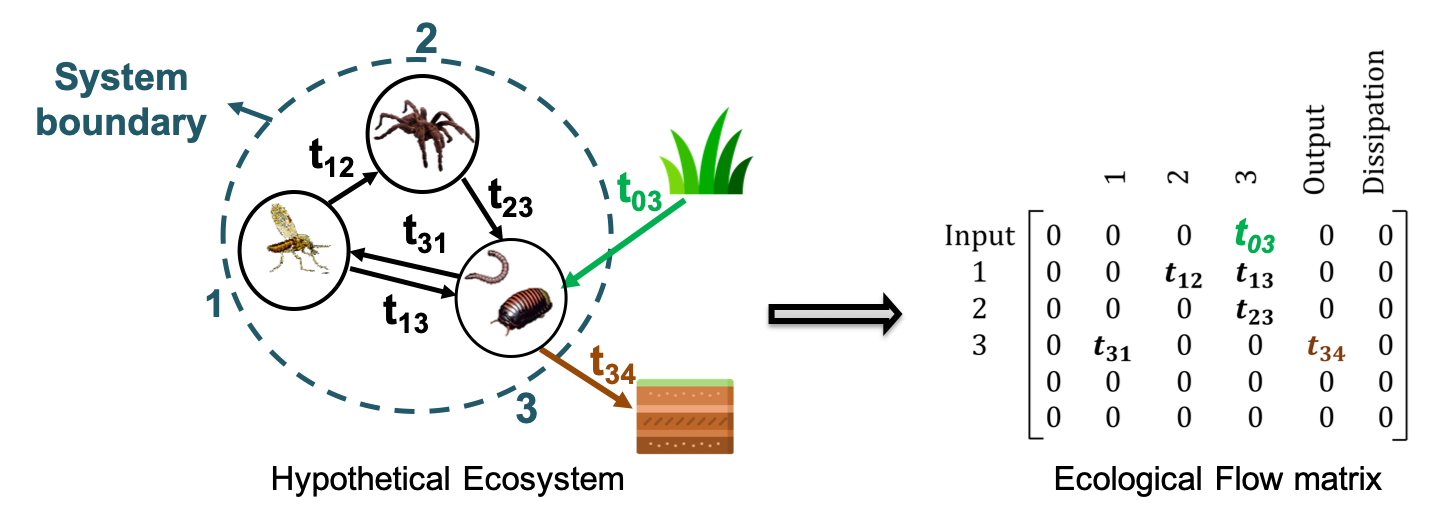}
  \caption{The conversion of a hypothetical ecosystem into Ecological Flow Matrix. Replicated from \cite{panyam2019bioT}.} 
  \label{hypotheticEFM}
  \vspace{-0.5cm}
\end{figure}

R\textsubscript{ECO} quantifies the robustness of a food web with specified system boundary. The input for calculating R\textsubscript{ECO} is the ecological flow matrix ([\textbf{T}]), which captures the energy interactions within and across the system boundary for a food web.
Fig \ref{hypotheticEFM} shows a hypothetical ecosystem and its conversion to [\textbf{T}]. The actors (species) that exchange energy based on a \textit{prey-predator} relationship are within the system boundary, and the energy providers, energy consumers, and energy dissipation are placed outside of the system boundary \cite{ulanowicz2012growth}.
Thus, [\textbf{T}] is a square ($N$+3) $\times$ ($N$+3) matrix containing flow magnitudes of transferred energy. \textit{N} is the number of actors inside the system boundary, and the extra three rows/columns represent the system inputs, useful system exports, and dissipation or system exports. It is essential to emphasize that the energy flows in and out of an actor and the system should be equal to maintain the \textit{law of conservation of energy}. 

%

The calculation of R\textsubscript{ECO} originates from the concept of \textbf{surprisal} and \textbf{indeterminacy} \cite{ulanowicz2009quantifying}, which are expressed as:
%
\begin{equation}\label{eq:surprisal}
s = -k \times log(p)	
\end{equation}
where $s$ is one's ``surprisal'' at observing an event that occurs with probability $p$, and $k$ is a positive scalar constant. Throughout this paper, the value of $k$ is one. 
%

\vspace{-0.1in}
\begin{equation}
\label{intermittency}
h_i = -k \times p_i \times log(p_i)	
\end{equation}
%
\noindent where $h_i$ is the indeterminacy of the event $i$. It is the product of the presence of an event $p_i$ and its absence $s_i$, which measures the potential for change with respect to an event $i$.

Thus, R\textsubscript{ECO} is able to quantify the ecosystems' robustness considering prospective events (disturbances) in the system with following metrics: \textbf{Total System Throughput} (\textit{TSTp}), \textbf{Ascendency} (\textit{ASC}), and \textbf{Development Capacity} (\textit{DC}).

The \textbf{Total System Throughput} (\textit{TSTp})
\cite{ulanowicz2009quantifying} is the sum of all flows in [\textbf{T}], capturing the system size: 
\vspace{-0.1in}
\begin{equation}
\label{eq:tstp}
\textit{TSTp} = \sum_{i=1}^{N+3}\sum_{j=1}^{N+3}T_{ij}
\end{equation}
\vspace{-0.1in}


The \textbf{Ascendency} (\textit{ASC}) measures the scaled mutual constraint for system size and flow organization that describes the process of ecosystems' growth and development \cite{Ulanowicz1980a}. It gives a dimensional version of network uncertainty as follow: 
\vspace{-0.1in}
\begin{equation} \label{eq:asc}
\textit{ASC} = -\textit{TSTp} \sum_{i=1}^{N+3} \sum_{j=1}^{N+3} \Bigg( \frac{T_{ij}}{\textit{TSTp}} log_2 \Bigg( \displaystyle \frac{T_{ij} \textit{TSTp}} {T_{i} T_{j}} \Bigg) \Bigg)
\end{equation}


\noindent where $T_{i}$ is the $\sum_{j=1}^{N+3}$ $T_{ij}$. The $\frac{T_{ij}}{TSTp}$ can be recognized as the probability of an event respect to the flows in the system. The $\frac{T_{ij}TSTp}{T_{i}T_{j}}$ is the conditional probability considering the knowledge of source (\textit{i}) and end (\textit{j}) nodes for a given flow (\textit{ij}) in the network. 
%

The \textbf{Development Capacity} (\textit{DC}) was introduced by Ulanowlcz in \cite{ulanowlcz1990symmetrical} as the upper bound of \textit{ASC} since there is a limit of ecosystems' growth and development. It is a food web's aggregate \textit{indeterminacy}, capturing the aggregated impacts (uncertainty) from all events (surprisals). 
\begin{equation} \label{eq:dc}
\textit{DC} = -\textit{TSTp} \sum_{i=1}^{N+3} \sum_{j=1}^{N+3} \Bigg(\displaystyle \frac{T_{ij}}{\textit{TSTp}} log_2 \Big(\displaystyle \frac{T_{ij}}{\textit{TSTp}}\Big)\Bigg)
\end{equation}
\vspace{-0.1in}


%

Then, the formulation of R\textsubscript{ECO} is as follows:

\begin{equation}\label{eq:r}
    \textit{\textit{\textit{R\textsubscript{ECO}}}} = -\bigg( \displaystyle \frac{\textit{ASC}}{\textit{DC}} \bigg) ln\bigg(\displaystyle \frac{\textit{ASC}}{\textit{DC}}\bigg)
\end{equation}

\noindent where the ratio of \textit{ASC} and \textit{DC} reflects the \textit{pathway efficiency} for a given network while its natural logarithm shows the network’s \textit{pathway redundancy} \cite{ulanowicz2009quantifying}.

The key feature of robust and resilient food webs is the unique balance of their networks' \textit{pathway efficiency} and \textit{pathway redundancy}, making the value of R\textsubscript{ECO} fall into the range of \textit{Window of Vitality} (around 0.3679) \cite{ulanowicz2009quantifying}.



\section{Power Systems' Ecological Flow Matrix}

The previous works \cite{panyam2019bioT, huangreco, huangreco2022} modeled real power flows as energy transfer and power grid components (buses and generators) as analogous to food web species to formulate [\textbf{T}] and calculate R\textsubscript{ECO} through equation (\ref{eq:tstp})-(\ref{eq:r}). However, power flows include both real and reactive power flows. Equation (\ref{pf})-(\ref{7l}) show the power flow and power balance equations in power systems for solving the power flow distribution problem. Reactive power flows are to support and stabilize the voltage of power systems. Besides, the apparent power, $S = P + iQ$, is used in power system operational limits assessment combining both real and reactive power in a complex number form.

\vspace{-0.1in}
\begin{equation}
P_{ij} = V_i^2[-G_{ij}] + 
V_i V_j [G_{ij} cos(\theta_{ij}) + B_{ij} sin(\theta_{ij})] 
\label{pf}
\end{equation}
\vspace{-0.1in} 
\begin{equation}
Q_{ij} = V_i^2[B_{ij}] + 
V_i V_j [G_{ij} sin(\theta_{ij}) - B_{ij} cos(\theta_{ij}) ] 
\label{rf}
\end{equation}
\vspace{-0.1in} 
\begin{equation}
P_{i} =P_{load_{i}}-P_{gen_{i}}= \sum_{j}P_{ij} 
\label{7k}
\end{equation}
\vspace{-0.1in} 
\begin{equation}
Q_{i} =Q_{load_{i}}-Q_{gen_{i}}= \sum_{j}Q_{ij} 
\label{7l}
\end{equation}
\vspace{-0.1in}



This paper extends the model of [\textbf{T}] in power systems with the consideration of \textit{reactive power flow} and \textit{apparent power flow} (magnitude). Unlike real power, the reactive power is associated with another power component, \textit{shunt capacitor}. It helps to regulate voltage, adjust power quality, and improve system reliability by generating or consuming reactive power locally. For reactive power, the generators can also either generate or consume reactive power by adjusting their setpoints for the system's stability. 

\begin{figure}[ht]
\centering
\includegraphics[width=0.98\linewidth]{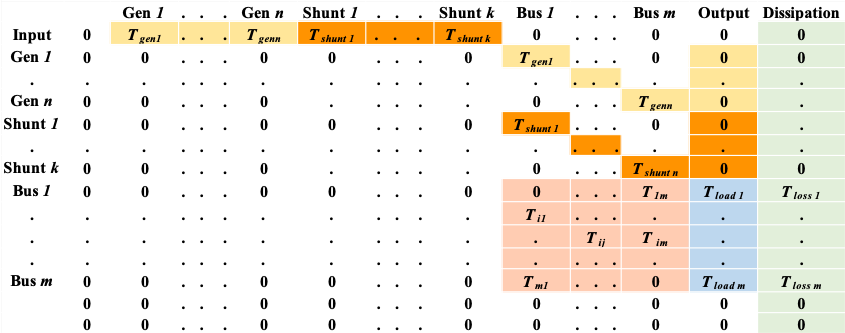}
  \caption{An extended \textit{Ecological Flow Matrix} [\textbf{T}] for any type of power flows in a grid with \textit{n} generators, \textit{k} shunt capacitors, and \textit{m} buses.} 
  \label{newflowmatrix}
  \vspace{-0.4cm}
\end{figure}

With above considerations, Fig. \ref{newflowmatrix} presents an extended model of [\textbf{T}] for any type of flow ($T_{ij}$) in power systems, which can be real power, reactive power, or apparent power, respectively. 
Shunt capacitor was not modeled in previous work as it didn't generate or consume real power, which is added in this model. 
It is treated as generator for reactive power and apparent power based [\textbf{T}]. The $T_{gen_{i}}$ is the flow from generator $i$. The $T_{shunt_{i}}$ is the flow from shunt $i$. Since generators and shunt capacitors can either generate or consume reactive power, the corresponding entries in [\textbf{T}] can either be \textit{Input} row or \textit{Output} column based on the flow direction. The $T_{load_{i}}$ and $T_{loss_{i}}$ are the power consumption and loss at Bus $i$, respectively. The $T_{ij}$ is the power flow at the corresponding branch from node $i$ to $j$. If there is no power flow interaction among buses, shunts, and generators, the entry is zero.


Except the flow type, this paper also looks into the local components' redundancy. In food webs, the energy transfer among species aggregates all animals of the same species to quantify the overall pathway efficiency and redundancy. However, in power systems, the same type components, like generators and shunt capacitors, connected to the same bus can have their own configurations and setpoints. Previous method aggregated the power flows if they are the same type component at the same bus. This may ignore the local network/component redundancy. Thus, with the extended model of [\textbf{T}], we construct two types of [\textbf{T}] that aggregates the power flows of the same type components at the same bus and splits the power flows to treat each component individually, respectively. With different [\textbf{T}], the value of R\textsubscript{ECO} will be different for capturing the system's inherent resilience. 

For the following cases studies, the calculation of R\textsubscript{ECO} considers different types of power flows using two approaches of dealing redundant local components, splitting and aggregating the flows, respectively. All above scenarios of modeling [\textbf{T}] and calculation of ecological metrics have been implemented in EasySimAuto (ESA) \cite{ESA}.


\section{Case Studies and Analyses}

In this paper, we analyze the R\textsubscript{ECO} considering above mentioned scenarios for two power systems cases, the IEEE 24 Bus RTS System and the Reduced GB Network with optimal power flow (OPF), \textit{N-1} security constrained optimal power flow (SCOPF), \textit{N-x} SCOPF, and the R\textsubscript{ECO} OPF in \cite{huangreco}. All the cases are available at \cite{BioCases}. To investigate the power flow distribution of all cases, Table \ref{FlowProperty} shows the \textit{Mean} and \textit{Standard Deviation (STD)} of all types power flow distribution. 


\begin{table}[hbt]
\caption{The Statistic Analyses of Power Flow Distribution}
\begin{adjustbox}{width=0.47\textwidth}
{\renewcommand{\arraystretch}{1.5}
\begin{tabular}{>{\centering}m{2.8cm}| >{\centering\arraybackslash}m{1.2cm}|>{\centering\arraybackslash}m{1.2cm}| >{\centering\arraybackslash}m{1.2cm}|>{\centering\arraybackslash}m{1.2cm}|>{\centering\arraybackslash}m{1.3cm}|>{\centering\arraybackslash}m{1.2cm}} 

\hline
\hline
    \textbf{Use Case} & Mean(pf) & STD(pf) & Mean(rf) & STD(rf) & Mean(MVA) & STD(MVA)   \\   

\hline

\hline

IEEE 24 Bus RTS       & 117.19          & 86.74            & 27.95 & 23.52 & 124.07  & 84.84  \\ \hline
IEEE 24 Bus RTS OPF           & 115.60          & 83.11             & 27.30 & 23.53  & 122.26 & 81.38 \\ \hline
IEEE 24 Bus RTS \textit{N-1} SCOPF    & 93.73         & 63.30             & 30.37 & 30.87 & 106.89 & 56.95  \\ \hline
IEEE 24 Bus RTS \textit{N-x} SCOPF    & 86.34           & 63.06              & 30.15 & 31.18 & 100.42 & 56.81 \\ \hline
IEEE 24 Bus RTS DC R\textsubscript{ECO} OPF   & 100.24          & 68.72             & 26.83 & 23.62 & 106.80  & 68.13  \\ \hline
IEEE 24 Bus RTS QCLS R\textsubscript{ECO} OPF  & 104.21          & 68.38            & 26.92 & 23.21 & 110.36 & 67.97 \\ \hline
\hline
Reduced GB Network                & 575.53         & 471.74             & 89.21 & 104.74 & 593.79 & 469.16 \\ \hline
Reduced GB Network OPF           & 555.97          & 444.76             & 90.19 & 104.55 & 574.24 & 442.98 \\ \hline
Reduced GB Network \textit{N-1} SCOPF &    522.69	&442.07	&85.65	&104.70	&540.13	&441.80 \\ \hline
Reduced GB Network \textit{N-x} SCOPF    & 436.39         & 392.80              & 83.66 & 103.39 & 453.72 & 395.68 \\ \hline
Reduced GB Network DC R\textsubscript{ECO} OPF  & 377.60          & 355.16             & 82.69 & 103.18 & 394.74 & 361.08 \\ \hline
Reduced GB Network QCLS R\textsubscript{ECO} OPF & 427.73          & 400.50            & 92.45 & 109.03 & 450.14 & 401.44 \\ \hline
\end{tabular}}

\end{adjustbox}
\label{FlowProperty}
\end{table}

\subsection{IEEE 24 Bus RTS System}

\begin{figure}[t!]
    \centering 
\begin{subfigure}[R\textsubscript{ECO} comparison with different flows and models]
  {\includegraphics[width=0.93\linewidth]{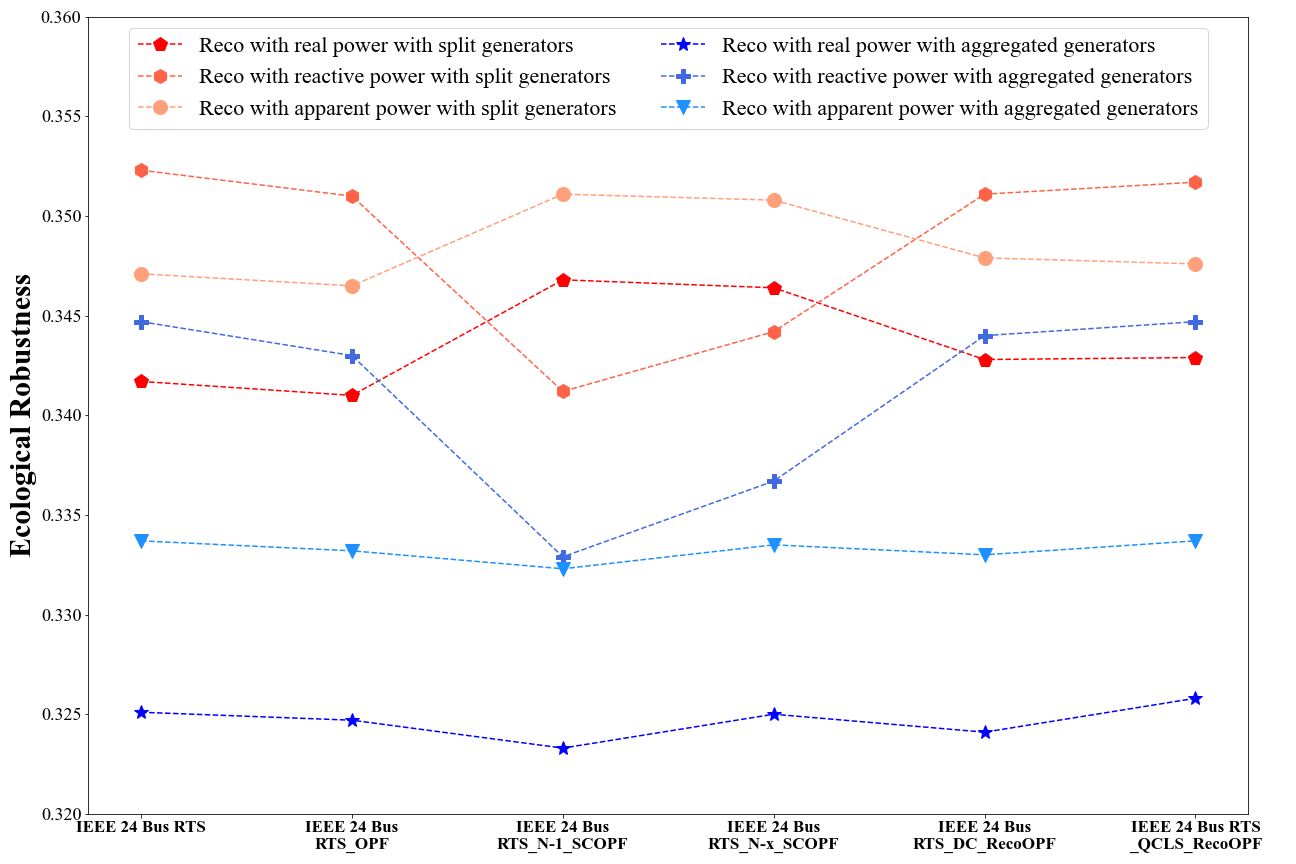}\label{24reco}}
\end{subfigure} 
\begin{subfigure}[Survivability Comparison among different optimal power flows. Replicated from \cite{huangreco}]
  {\includegraphics[width=0.93\linewidth]{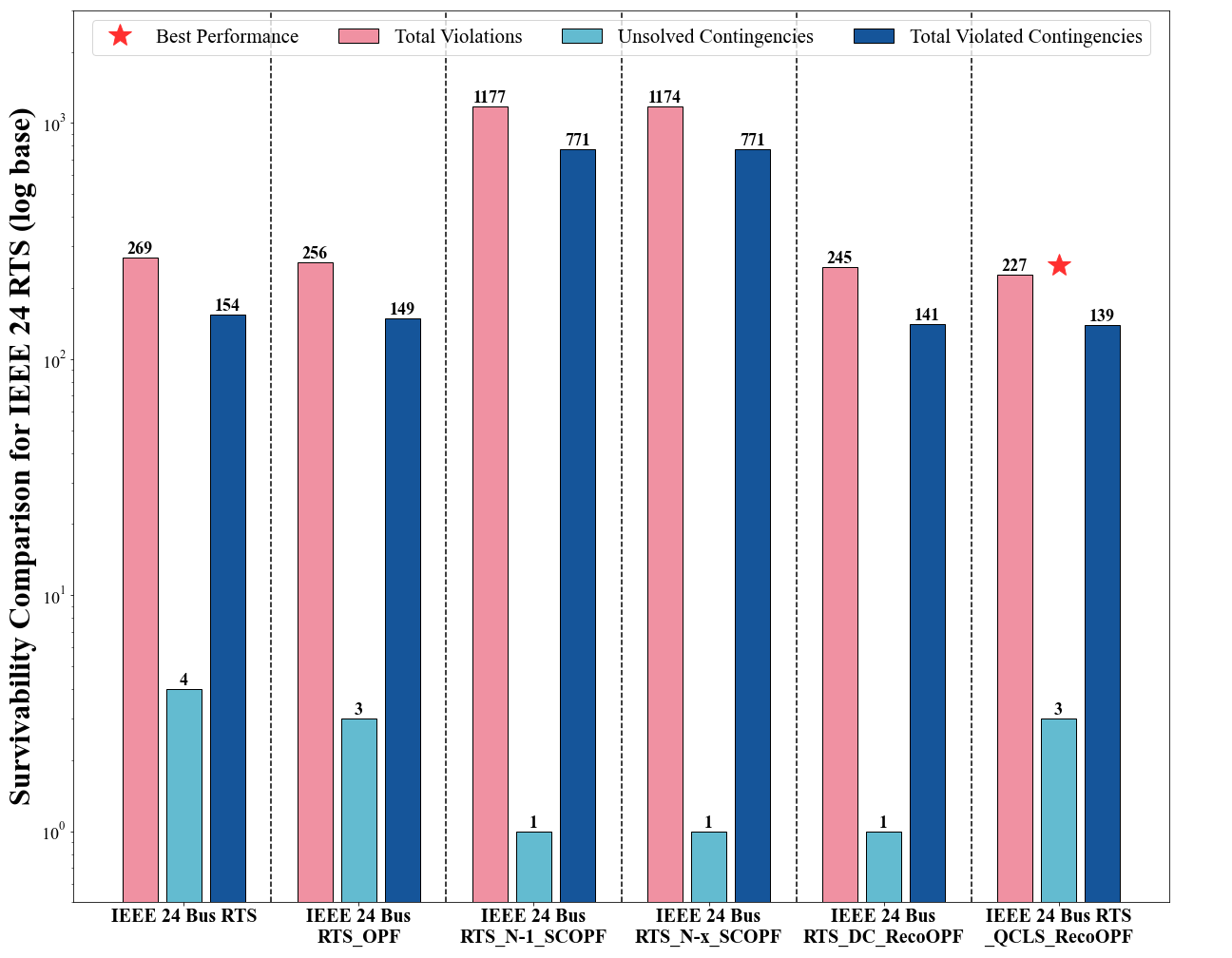}\label{24survivability}}
\end{subfigure} 
\caption{R\textsubscript{ECO} analyses for IEEE 24 Bus RTS System under different optimal power flow algorithms}
\label{ieee24}
\vspace{-0.4cm}
\end{figure}

For the IEEE 24 Bus RTS system, there is one shunt capacitor in the system and several buses have connected multiple generators (2 to 4). Fig. \ref{24reco} shows the value of R\textsubscript{ECO} with real, reactive, and apparent power considering aggregated generators and split generators, respectively. To better correlate R\textsubscript{ECO} with resilience, Fig. \ref{24survivability} shows each optimal power flow algorithm's survivability under \textit{N-x} contingencies (removing \textit{x} components from the system), quantified as the number of violations, unsolved contingencies, and violated contingencies. 

First of all, Fig. \ref{24reco} shows that when the redundancy of generators is taken into consideration, the corresponding value R\textsubscript{ECO} is higher than the ones that ignore generators' redundancy. For the same type power flow, all red lines are higher than the blue lines. The local redundancy on power system components can contribute to improvement on R\textsubscript{ECO}. 
It suggests an application of using R\textsubscript{ECO} to guide the placement of generator units, such as distributed energy resources, for improving the overall resilience. Secondly, under the same network structure (split redundant generators' flows or aggregate them), the value of reactive power flow based R\textsubscript{ECO} is the highest (except for the \textit{N-1} SCOPF and \textit{N-x} SCOPF); the real power flow based R\textsubscript{ECO} is the lowest; and the apparent power flow based R\textsubscript{ECO} is in between. It shows that the distribution of reactive power is the most robust than other flows even their network structure is the same. Even though apparent power considers both real and reactive power, the magnitude of real power is bigger than reactive power and the direction of apparent power flow is assumed to follow real power flow. Thus, the trend of apparent power based R\textsubscript{ECO} is close to the real power based R\textsubscript{ECO}.

The discrepancies of R\textsubscript{ECO} on \textit{N-1} and \textit{N-x} SCOPF cases can be well explained by the survivability under \textit{N-x} contingencies and the statistical analyses of flow distribution. Fig. \ref{24survivability} shows that the R\textsubscript{ECO} OPF using the quadratic-convex relaxed alternating current (QCLS) model has the best performance 
with minimum violations and unsolved contingencies while the \textit{N-1} and \textit{N-x} SCOPF are the worst. Only the reactive power flow based R\textsubscript{ECO} reflects this pattern, and the trend on R\textsubscript{ECO} is fairly close to the trend of survivability comparison. As most of the violations are voltage related, which is associated with reactive power flow distribution, it makes sense that reactive power flow based R\textsubscript{ECO} can better capture the system's inherent ability of absorbing disturbances. This finding was not captured in previous work. With the statistical evaluation of all types power flows, the reactive power is more equally distributed in R\textsubscript{ECO} OPF cases than the SCOPF cases. This also justifies that the homogeneity on flows contributes to their ability of dealing with cascading failures \cite{huangreco}. 

\subsection{Reduced Great Britain Network}
\begin{figure}[t!]
    \centering 
\begin{subfigure}[R\textsubscript{ECO} comparison with different flows and models]
  {\includegraphics[width=0.93\linewidth]{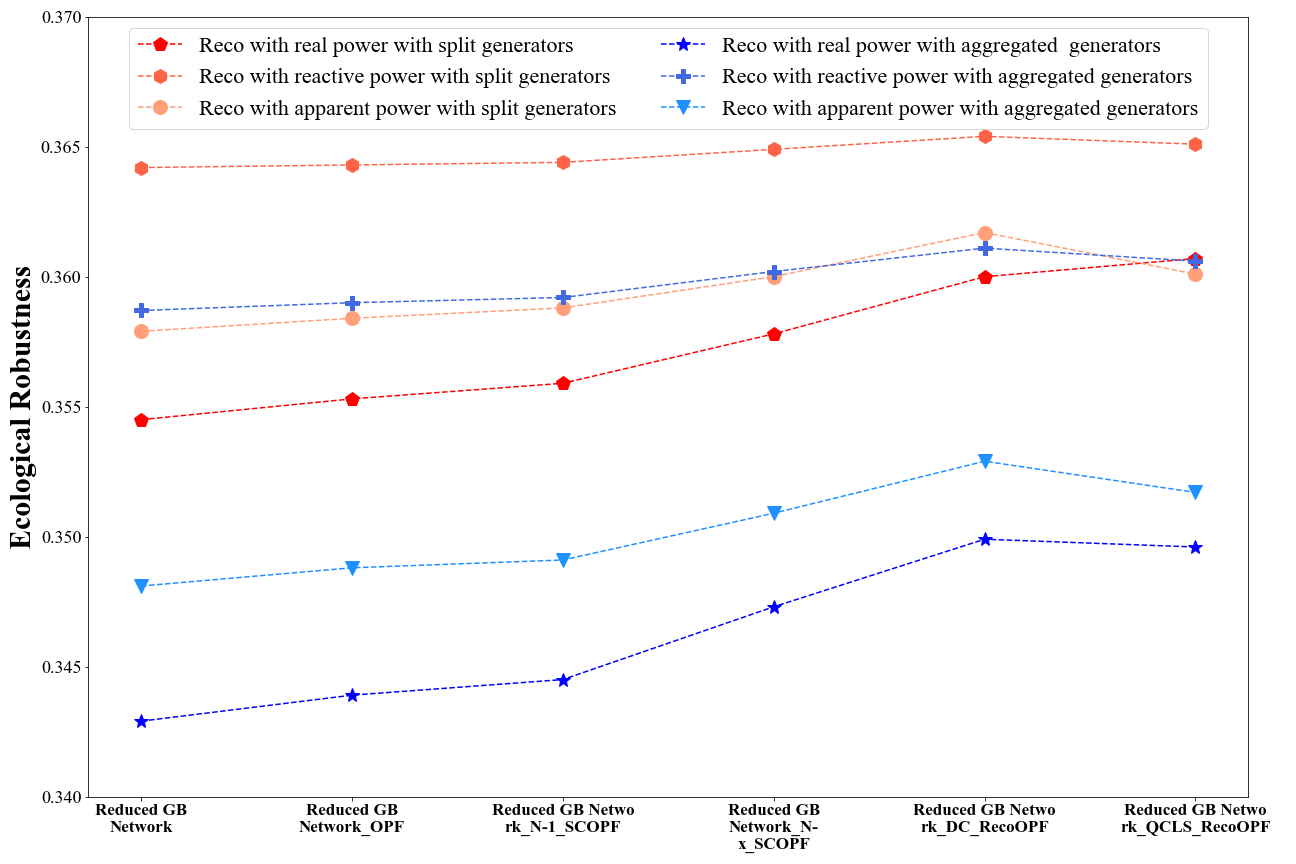}\label{GBreco}}
\end{subfigure} 
\begin{subfigure}[Survivability comparison among different optimal power flows. Replicated from \cite{huangreco}]
  {\includegraphics[width=0.93\linewidth]{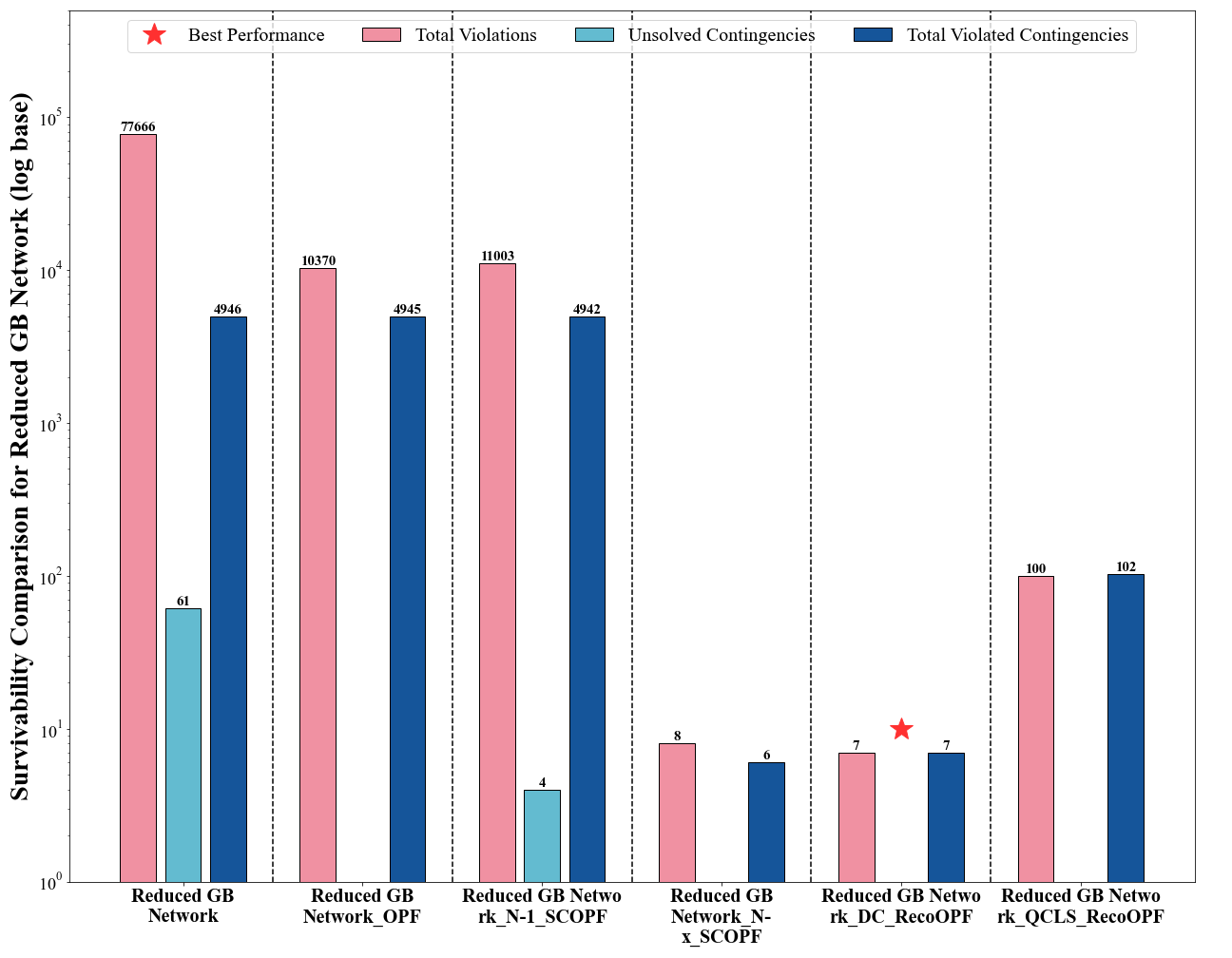}\label{GBsurvivability}}
\end{subfigure} 
\caption{R\textsubscript{ECO} analyses for Reduced Great Britain (GB) Network under different optimal power flow algorithms}
\label{fig:images}
\vspace{-0.4cm}
\end{figure}

For the Reduced GB Network, there is no shunt capacitor in the system but several buses have connected with multiple generators (2 to 5). Similarly, Fig \ref{GBreco} shows the value of R\textsubscript{ECO} with real, reactive, and apparent power flows considering aggregated generators and split generators, respectively; and Fig \ref{GBsurvivability} shows each cases' survivability under \textit{N-x} contingencies. 

The results are similar to the IEEE 24 Bus RTS System. Firstly, the value of R\textsubscript{ECO} is higher when the redundancy of generators is taken into consideration. Secondly, with the same structure of [\textbf{T}], the reactive power based R\textsubscript{ECO} is the highest, followed by the apparent power based and real power based. Thirdly, for this case, except the real power based R\textsubscript{ECO}, the trend of R\textsubscript{ECO} with other flows fairly correlates with the survivability and the statistical analyses on the flow distribution. The highest R\textsubscript{ECO} has the best performance under \textit{N-x} contingencies with the most equally distributed flows. Less resilient cases have worse R\textsubscript{ECO} and less equally distributed power flows.


\section{Conclusion and Future Work}

This paper has presented an extended model to map power systems with ecosystems for applying R\textsubscript{ECO} to evaluate power systems' robustness and resilience. From the case studies, we can draw the following conclusions. Firstly, the reactive power based R\textsubscript{ECO} can better capture the systems' resilience and survivability as voltage violations are predominate under \textit{N-x} contingency analyses. There is a significant potential of using both real and reactive power based R\textsubscript{ECO} as a resilience index to evaluate and optimize the system's inherent ability of absorbing disturbances. Secondly, it has been observed that the components' redundancy can contribute the improvement of R\textsubscript{ECO}. \textcolor{black}{This suggests a new direction of using R\textsubscript{ECO} to strategically place generator units, such as distributed energy resources and renewable energy sources. The stochastic nature of renewable energy sources is reflected in their outputs (real and reactive power), which can be taken into account for an R\textsubscript{ECO}-oriented stochastic optimization. That will strategically add local redundancy for improving the overall robustness and resilience of the system.}

\section*{Acknowledgment}
The authors would like to acknowledge the National Science Foundation under Grant 1916142 and 2039716, the US Department of Energy Cybersecurity for Energy Delivery Systems program under award DE-OE0000895, and a grant from the C3.ai Digital Transformation Institute for their support of this work. 

\bibliographystyle{IEEEtran}

\bibliography{hreference.bib}

\end{document}